\documentclass[aps,prd,showpacs,twocolumn,superscriptaddress,nofootinbib,tightenlines,preprintnumbers]{revtex4-2} 

\makeatletter
\def\p@subsection{}
\makeatother

\usepackage{graphicx,amsmath,amssymb,lipsum,bm}

\usepackage[usenames,dvipsnames]{xcolor}
\definecolor{xlinkcolor}{rgb}{0.7752941176470588, 0.22078431372549023, 0.2262745098039215}

\usepackage[colorlinks=true,citecolor=xlinkcolor,linkcolor=xlinkcolor,urlcolor=xlinkcolor, backref=false,pdfborder={0 0 0}]{hyperref}

\usepackage[normalem]{ulem}
\usepackage[utf8]{inputenc} 
\usepackage{mathtools}
\usepackage{aas_macros}
\usepackage{float}

\usepackage{booktabs}

\def\beqn#1{\begin{equation}\label{#1}}
\def\eeqn{\end{equation}}

\def\beqa#1{\begin{eqnarray}\label{#1}}
\def\eeqa{\end{eqnarray}}
\def\beq{\begin{eqnarray}}
\def\eeq{\end{eqnarray}}
\let\vec\mathbf

\usepackage{ulem}
\usepackage{diagbox}

\usepackage[dvipsnames]{xcolor}
\usepackage{mathrsfs}

\newcommand{\be}{\begin{equation}}
\newcommand{\ee}{\end{equation}}

\newcommand{\bseq}{\begin{subequations}}
\newcommand{\eseq}{\end{subequations}}

\renewcommand{\ln}{\mathop{\rm ln}\nolimits}

\newcommand{\ld}{\Lambda{\rm CDM}}

\newcommand{\wa}{w_0w_a{\rm CDM}}

\def\gsim{\raise0.3ex\hbox{$\;>$\kern-0.75em\raise-1.1ex\hbox{$\sim\;$}}}
\def\lsim{\raise0.3ex\hbox{$\;<$\kern-0.75em\raise-1.1ex\hbox{$\sim\;$}}}

\def\beqn#1{\begin{equation}\label{#1}}
\def\eeqn{\end{equation}}

\def\beqa#1{\begin{eqnarray}\label{#1}}
\def\eeqa{\end{eqnarray}}

\def\hMpc{\,h{\text{Mpc}}^{-1}}
\def\Mpch{h^{-1}{\text{Mpc}}}

\def\Z2{$\mathcal{Z_2}$}

\defcitealias{desi1}{Paper 1}
\defcitealias{desi2}{Paper 2}
\defcitealias{desi3}{Paper 3}
\defcitealias{desi4}{Paper 4}
\defcitealias{desi5}{Paper 5}

\usepackage{xspace}

\newcommand {\ignore}[1]{}

\renewcommand{\arraystretch}{1.3}

\begin{document}

\preprint{MIT-CTP/6122}

\title{Constraints on Primordial Oscillatory Features from the Power Spectrum and Bispectrum of Redshift-Space 
Galaxy Clustering in DESI DR1 
}

\author{Shi-Fan Chen}
\email{sc5888@columbia.edu}
\affiliation{Department of Physics, Columbia University, New York, NY 10027, USA}
\affiliation{NASA Hubble Fellowship Program, Einstein Fellow}

\author{Anton~Chudaykin}
\email{anton.chudaykin@unige.ch}
\affiliation{D\'epartement de Physique Th\'eorique and Center for Astroparticle Physics,\\
Universit\'e de Gen\`eve, 24 quai Ernest  Ansermet, 1211 Gen\`eve 4, Switzerland}

\author{Mikhail M.~Ivanov}
\email{ivanov99@mit.edu}
\affiliation{Center for Theoretical Physics -- a Leinweber Institute, Massachusetts Institute of Technology, 
Cambridge, MA 02139, USA} 
 \affiliation{The NSF AI Institute for Artificial Intelligence and Fundamental Interactions, Cambridge, MA 02139, USA}

\author{Oliver~H.\,E.~Philcox}
\email{ohep2@cantab.ac.uk}
\affiliation{Leinweber Institute for Theoretical Physics at Stanford, 382 Via Pueblo, Stanford, CA 94305, USA}
\affiliation{Kavli Institute for Particle Astrophysics and Cosmology, 382 Via Pueblo, Stanford, CA 94305, USA}

\author{Mario Ballardini}
\email{mario.ballardini@unife.it}
\affiliation{Dipartimento di Fisica e Scienze della Terra, Universit`a degli Studi di Ferrara, via Giuseppe
Saragat 1, 44122 Ferrara, Italy}
\affiliation{INFN, Sezione di Ferrara, via Giuseppe Saragat 1, 44122 Ferrara, Italy}
\affiliation{
INAF-Osservatorio di Astrofisica e Scienza dello Spazio di Bologna, Via Piero Gobetti 93/3,
40129 Bologna, Italy}
\affiliation{Department of Physics \& Astronomy, University of the Western Cape, Cape Town 7535,
South Africa}

\begin{abstract}
\noindent We constrain oscillatory features in the primordial power spectrum
using the public DESI DR1 galaxy and quasar clustering data.
Our analysis is based on a custom pipeline for the redshift-space 
power spectrum and bispectrum of the six main DESI DR1 tracers,
modeled with the Lagrangian effective field theory,
which provides 
a principled 
resummation of the non-linear damping of the features. We consider oscillations that are linear and logarithmic in wavenumber,
and, for the first time, consistently include the galaxy bispectrum in the search. 
We also study the dependence of the results on the background cosmology, 
which is important given the tension between the DESI BAO and \textit{Planck} data within the baseline cosmological model. Marginalizing over $\Lambda$CDM parameters, we bound the amplitude of 
linear and logarithmic oscillations to $A_{\rm lin}<0.033$ and $A_{\rm log}<0.035$ respectively (at 95\%~CL) across the frequency ranges $\omega_{\rm lin}^{\rm phys}\in(20,800)~{\rm Mpc}$ 
and $\omega_{\rm log}\in(2.5,80)$. The constraints reach percent-level precision away from the frequencies degenerate with the baryon acoustic oscillations (BAO).
Fixing to the \textit{Planck} 2018 cosmology, the bounds tighten to
$A_{\rm lin}<0.024$ and $A_{\rm log}<0.031$, but in this case 
we also find a $\simeq 3\sigma$ global preference for a logarithmic 
feature with $\omega_{\rm log}\simeq 15$, dominated by the LRG2 sample, whose 
local wavelength
matches the apparent 
BAO scale 
at wavenumbers best constrained by DESI.
Since the preference for this feature disappears once the cosmological parameters are varied,
we expect that it is a manifestation of the known distance-scale 
tension between the DESI and \textit{Planck} data rather than a primordial signal. Our analysis
thus emphasizes the importance
of cosmology marginalization
in searches for primordial 
features. 
\end{abstract}

\maketitle


\section{Introduction}
\label{sec:intro}

\noindent Cosmic inflation is the leading 
theoretical framework for explaining the observed properties of  
cosmological fluctuations, namely their
adiabatic nature, Gaussian statistics, 
and the nearly scale invariant 
spectrum of primordial 
density fluctuations. 
The simplest inflationary models
predict only relatively weak modifications
of the initial power spectrum, captured by the tilt and the running of the spectral index. More broadly, a class of inflationary models based on 
ideas of  axion monodromy inflation~\cite{Silverstein:2008sg}
predicts the presence of oscillating 
features in the primordial density
fluctuations, 
which reflect the modulations of
the inflaton potential by harmonic
oscillations~\cite{Pahud:2008ae,Flauger:2009ab,Flauger:2010ja}.
In addition, oscillations in
the 
primordial power spectrum and
the 
statistics of the matter density tracers
can be generated by the exchange of massive particles, which is known as the cosmological collider scenario~\cite{Ghosh:2014kba,Chen:2009we,Chen:2009zp,Chen:2010xka,Byrnes:2010xd,Achucarro:2010da,Baumann:2011nk,Chen:2012ge,Noumi:2012vr,Assassi:2012zq,Arkani-Hamed:2015bza,Lee:2016vti,Kehagias:2017cym,An:2017hlx,Kumar:2017ecc,Bordin:2018pca,Kumar:2018jxz,Goon:2018fyu,Goon:2018fyu,Hook:2019zxa,Kumar:2019ebj,Liu:2019fag,Wang:2019gbi,Bodas:2020yho,Aoki:2020zbj,Lu:2021wxu,Pinol:2021aun,Reece:2022soh,MoradinezhadDizgah:2017szk,MoradinezhadDizgah:2018ssw,MoradinezhadDizgah:2019xun,Jense:2026haw}.\footnote{In general, cosmological collider models produce oscillations
in the primordial bispectrum, which then are imprinted in the galaxy power spectrum by the scale-dependent bias. 
There are, however, cosmological-collider-like  
models with oscillating massive fields that generate resonant features directly in the primordial power spectrum~\cite{Chen:2014cwa,Braglia:2021rej}. }

Oscillatory primordial features
have been constrained 
using CMB and LSS data in the past~\cite{Martin:2003sg,Meerburg:2011gd,Meerburg:2013dla,Miranda:2013wxa,Planck:2015sxf,Planck:2018jri,Zeng:2018ufm,Beutler:2019ojk,Ballardini:2022wzu,Antony:2024vrx,Calderon:2025xod,Raffaelli:2026bhk,Nerval:2026iev}.
The latter probe is believed 
to be more promising in the long run thanks to its access to the 
three-dimensional distribution
of cosmological modes~\cite{Chen:2016vvw,Ballardini:2016hpi,Chen:2016zuu,Xu:2016kwz,Palma:2017wxu,Slosar:2019gvt}. Another attractive feature of searches for
primordial oscillations 
in the LSS data is their stability with respect to non-linear effects provided that the typical correlation length of the oscillations is greater than $\sim 10~\Mpch$. 
While non-linear evolution
damps the amplitude of the primordial features 
in the LSS statistics~\cite{Vasudevan:2019ewf,Beutler:2019ojk,Chen:2020ckc,Ballardini:2024dto}, 
the shift of their frequency 
is quite minor because the 
non-linear 
coupling of these 
features to 
long-wavelength modes is controlled
by the large-scale mass variance, which happens to be sufficiently small in our Universe~\cite{Padmanabhan:2008dd,Sherwin:2012nh,Ivanov:2022mrd,Chen2024}.
The same argument ensures the
robustness of another 
important feature in the 
LSS statistics: baryon acoustic oscillations (BAO)~\cite{Smith:2006ne,Crocce:2007dt,Padmanabhan:2008dd,Sherwin:2012nh,Blas:2016sfa}. 

In this work we update the constraints on oscillatory primordial features using the 
public DESI DR1 galaxy and quasar clustering data \citep{DESI:2025fxa}. In addition to using the state-of-the-art LSS dataset, our work features several other improvements over previous studies. 
First, we consistently 
include the galaxy bispectrum
in our analysis. 
Second, we marginalize our constraints over background cosmological models, which is especially important given the $\Lambda$CDM tension between DESI and \textit{Planck} CMB data, which has been  interpreted as 
evidence for evolving dark energy~\cite{DESI:2025gwf,desi2}. 

Our paper is structured as follows. In Section~\ref{sec:data} we describe the data, with the feature templates and the Lagrangian perturbation theory model introduced in Section~\ref{sec:theory}. Section~\ref{sec:ppt} describes details of our pipeline, including model parameters and priors, as well as the look-elsewhere correction for linear and logarithmic primordial features, before we present the baseline cosmology fits in Section~\ref{sec:cosmo} and the DESI constraints on features in Section~\ref{sec:results}. We conclude in Section~\ref{sec:conclusions}. Appendix~\ref{app:pt_challenge} validates the pipeline on the PT Challenge simulation.

\section{Data}
\label{sec:data}

\noindent Our constraints use custom measurements of the DESI DR1 power spectrum and bispectrum, obtained using the pipeline described in Refs.~\citep{desi1,desi2,desi3}. 
Following \citep{DESI:2024aax}, we include six samples: BGS, LRG1, LRG2, LRG3, ELG2 and QSO, encompassing bright galaxies (BGS), luminous red galaxies (LRG), emission line galaxies (ELG) and quasars (QSO), across a total redshift range of $0.1<z<2.1$, and carefully account for a number of observational complications, including masking, integral constraints, fiber collisions, and the masked Gaussian covariance. In addition to the power spectrum multipoles, $P_\ell$, and bispectrum monopole, $B_0$, used in previous works, we include the real-space power spectrum proxy, $Q_0$, defined in \cite{Ivanov:2021fbu}. This statistic, formed from the power spectrum multipoles as $Q_0 = P_0-\tfrac12P_2+\tfrac38P_4\approx P(k,\mu=0)$, suppresses fingers-of-God contributions, 
such that the one-loop perturbative description is accurate down to smaller scales. While the restriction to transverse modes would na\"ively lead to increased sensitivity to fiber collision effects, we explicitly subtract close pairs from our estimator, removing such artifacts \citep{desi1}. Here, we use the power spectrum multipoles up to $0.2\hMpc$, with $Q_0$ used in the range 
$0.2\hMpc<k<0.4\hMpc$ (itself assembled from new high-resolution measurements of the power spectrum).

To optimize our analysis for the search
of oscillating features, we use a finer Fourier space binning than in previous works, with $\Delta k=0.005\hMpc$ except for QSOs where we adopt $\Delta k=0.0025\hMpc$. 
Our bispectrum binning is
$\Delta k=0.01\hMpc$,
the same as in~\cite{desi1}, and we use $k_{\rm min}=0.02\hMpc$
for all statistics. 
We use conservative priors on EFT parameters matching those from \citep{desi2} (see \citep{desi5} for an extension with simulation-based priors).

For the bispectrum, we use the tree-level likelihood from \citep{desi1,desi2}. 
While the one-loop bispectrum likelihood
for DESI DR1 is also available~(see \citep{desi3,desi4,Smith:2026bmb}), 
we do not include small-scale
bispectrum data in this work because it requires a more general non-linear 
damping that goes beyond 
the current modeling well tested 
for the BAO only~\cite{Ivanov:2018gjr,Philcox:2022frc,Bakx:2025pop}. 

In order to
further constrain
the background expansion, we make use of two additional datasets: the DESI DR2 BAO likelihood~\cite{DESI:2025zgx} (which \citep{desi1} showed to be negligibly correlated with the DR1 full-shape on the scales of interest), and the uncalibrated Pantheon+ supernovae data \citep{Brout:2022vxf} (SNe).

\section{Theory}
\label{sec:theory}

\noindent We constrain the following 
modification
of the scalar power spectrum:
\begin{align}
    \frac{P_{\zeta}(k)}{P_\zeta^{\rm fid}(k)} - 1 &= A_X \sin( f_X(k) + \phi ) \nonumber \\
    &= A^X_{\rm sin} \sin f_X(k) + A^X_{\rm cos} \cos f_X(k)
\end{align}
where $f_{\rm lin}(k) = \omega_{\rm lin} \,k/k_\ast$ for linear oscillations and $f_{\rm log}(k) = \omega_{\rm log} \log(k/k_\ast)$ for logarithmic oscillations, where $k_\ast = 0.05\ \text{Mpc}^{-1}$ is the pivot scale typically defined for the CMB. 
In the linear case, we will also report the results in terms of the physical frequency $\omega^{\rm phys}_{\rm lin} = \omega_{\rm lin}/k_\ast$.

The fiducial scalar power spectrum $P^{\rm fid}_\zeta$ follows the standard parametrization defined by its amplitude $A_s$ at the pivot scale and spectral index $n_s$, and the linear matter power spectrum is obtained by multiplying by the transfer function, which we will compute both within $\Lambda$CDM and dynamical dark energy cosmologies. In practice, we sample the amplitudes of the sine and cosine modes $A^X_{\rm sin,\,cos}$ rather than the total amplitude $A_X = \sqrt{(A^X_{\rm sin})^2 + (A^X_{\rm cos})^2}$ and phase $\phi$, since the former parameters decorrelate in the limit that the phase $f_X(k)$ varies by $\gg 1$ over the scales analyzed, while $A,\phi$ feature strong degeneracies and posteriors that jump over the boundaries of $\phi \in [0, 2\pi)$. It is straightforward to convert between the two parametrizations in post-processing.

In this paper, we will use the Lagrangian effective field theory of large-scale structure (LPT) to compute predictions for galaxy clustering given the background cosmology and linear power spectra described above \cite{Matsubara:2007wj,Matsubara2008,Carlson:2012bu,Porto:2013qua,Vlah:2015sea,Vlah:2016bcl,Chen:2020zjt,Chen:2020fxs}.
When searching for primordial features in galaxy clustering datasets, it is critical to correctly account for the smearing of oscillatory components in the power spectrum by the long-wavelength (IR) motions of galaxies. Since the typical size, $\Sigma_X$, of these motions on the scale $r_X$ of the feature (\textit{i.e.}\ $r_{\rm lin} = \omega_{\rm lin}^{\rm phys}$ and $r_{\rm log} = \omega_{\rm log}/k$) can be large compared to the observed wavelength for a broad range of frequencies, their contributions need to be properly resummed to keep loop corrections well-behaved. This procedure, known as IR resummation, has been extensively studied in the context of both Eulerian and Lagrangian EFTs of large-scale structure \cite{Carlson:2012bu,Matsubara:2007wj,Crocce:2007dt,Senatore:2014via,Baldauf:2015xfa,Vlah:2015sea,Blas:2015qsi,Blas:2016sfa,Vlah:2015zda,Senatore:2017pbn,Ivanov:2018gjr,Vasudevan:2019ewf,Beutler:2019ojk,Chen:2020ckc,Chen:2024pyp}. 

One advantage of working in Lagrangian space is that resumming the nonlinear damping of oscillatory features occurs ``out-of-the-box'' (\textit{i.e.}\ without explicitly separating smooth and oscillatory components within the linear power spectrum) by keeping the effects of long-wavelength galaxy pairwise displacements exponentiated via the cumulant theorem. Briefly, the redshift-space galaxy power spectrum can be written in the form
\begin{equation}
    P_{g,s}(\vec k) = \int d^3 \vec q \ e^{i \vec k \cdot \vec q - \frac12 k_i k_j A^{s, \rm lin}_{ij}(\vec q)} \ F_{g,s}(\vec q)
    \label{eqn:lpt}
\end{equation}
where $A^{s, \rm lin}_{ij}(\vec q) = \langle (\Psi^s(\vec q) - \Psi^s(\vec 0) )_i (\Psi^s(\vec q) - \Psi^s(\vec 0) )_j\rangle^{\rm lin}$ is the mean variance of the redshift-space pairwise displacements of galaxies separated by $\vec q$ under the Zel'dovich approximation. Here
\begin{equation}
    \tilde{\Psi}^{s, \rm lin}_i(\vec k) = \left( \frac{i (k_i + f k \mu \hat{n}_i)}{k^2} \right) \tilde{\delta}_{\rm lin}(\vec k)
\end{equation}
and $F_{g,s}(\vec q)$ represents the nonlinear contributions to galaxy clustering excluding the resummed long-wavelength (Zel'dovich) contributions. An oscillatory feature in Fourier space picks up this variance at its localized scale in Lagrangian space at separations $|\vec q| = r_X$, \textit{i.e.}\ $\exp(-k_i k_j A^{s,\rm lin}_{ij}(\vec q)) \rightarrow \exp(-k^2 \Sigma^2(r_X)/2)$
leading to an apparent Gaussian damping of the feature \cite{Vlah:2015zda,Blas:2016sfa,Vasudevan:2019ewf,Chen:2020ckc,Chen:2024pyp}. In the Eulerian approach, this damping exponent is explicitly applied to the separated-out oscillatory component of the power spectrum.

Techniques to efficiently compute integrals of the form Eq.~\eqref{eqn:lpt} have been developed in \cite{Vlah15,Vlah:2018ygt,Chen19,Chen:2020zjt}. A particular feature of LPT and Eq.~\ref{eqn:lpt} is that the IR-resummed predictions must be computed separately for each wavenumber $k$, making calculations more computationally expensive than their Eulerian counterparts. For most practical analyses this issue is obviated because these computations are nonetheless faster than the Boltzmann codes needed to generate the linear power spectrum, and because the calculations can be sped up via emulators (e.g. \cite{DeRose2022}). However, a particularly expensive calculation, still subject to these constraints, is the galaxy bispectrum. The calculation of the bispectrum in LPT was developed in Ref.~\cite{Chen:2024pyp} using a generalization of the techniques utilized in power spectrum calculations; despite these advances, the numerical evaluation of the bispectrum remains prohibitively slow, owing to the need to integrate over IR kernels depending on internal momenta even at leading order. However, the authors of that work showed (see also \cite{Blas:2016sfa}) that the full LPT calculation can be well approximated by substituting 
\begin{equation}
    P_{\rm lin}^{\rm res}(\vec k) = \int d^3 \vec q \ e^{i \vec k \cdot \vec q - \frac12 k_i k_j A^{\rm s, lin}_{ij}(\vec q)} \ \xi_{\rm lin}(\vec q) 
    \label{eqn:plin-res}
\end{equation}
for the linear power spectrum input into tree-level and one-loop bispectrum calculations. In this work we will adopt this approximation to compute the large-scale galaxy bispectrum at tree level in the Eulerian EFT nomenclature. We describe the free parameters and functional form of the power spectrum and bispectrum in LPT in the next section.

\section{Parameters and Priors}
\label{sec:ppt}

\subsection{Model Details and Parameters}

\noindent We sample the $\Lambda$CDM parameters $\ln(10^{10} A_s), H_0, \omega_{cdm}$ with uninformative, uniform priors. The spectral tilt and baryon density are less well constrained by galaxy clustering, thus we sample them using Gaussian priors $n_s \sim \mathcal{N}(0.9649, 0.042^2)$, corresponding to a $10 \times$ the Planck constraint~\cite{Planck2018}, and $\omega_b \sim \mathcal{N}(0.02218, 0.00055^2)$ based on Big-Bang Nucleosynthesis \cite{Schoneberg:2024ifp,Burns:2023sgx}, where $\mathcal{N}(\mu,\sigma^2$) denotes a Gaussian distribution with mean $\mu$ and variance $\sigma^2$. When extending the analysis to dynamical dark energy cosmologies, we sample the equation of state parameters $w_0 \sim \mathcal{U}[-3,1]$ and $w_a \sim \mathcal{U}[-3,2]$, where $\mathcal{U}[a,b]$ denotes a uniform prior over the specified range. We additionally impose the matter domination constraint $w_0 + w_a < 0$. In all cases, we fix the neutrino mass to the minimal value: $M_\nu = 0.06\ \text{eV}$.

For our inflationary features constraints we adopt uniform logarithmic priors in $1.5 < \omega_{\rm lin} < 40$, corresponding to physical scales from $30-800\ \text{Mpc}$, and $2.5 < \omega_{\rm log} < 80$. These bounds are determined by a combination of the physical resolution of our data --- including the fitting range $k_{\rm min}$ and $k_{\rm max}$, set by the survey size and perturbativity, respectively, and the bin width $\Delta k$, effectively set by the survey window function --- as well as the convergence of the IR resummation in the case of logarithmic features, which become prohibitively slow to compute on a $k$-by-$k$ basis at high $\omega_{\rm log}$ and low $k$, where the corresponding physical frequency becomes quite fast. We sample $A^X_{\rm sin, cos}$ with uniform priors $\mathcal{U}[-0.5, 0.5]$. This corresponds to a uniform prior in the phase $\phi \sim \mathcal{U}[0,2\pi]$ but a non-uniform prior on the amplitude $P(A_{\rm lin}) \propto A_{\rm lin}$ owing to the increased parameter volume in polar coordinates at larger radii; in order to convert to bounds on $A_{\rm lin}$, we post-process our chains to remove this weighting.

We compute predictions for the resummed galaxy power spectrum as in Eq.~\ref{eqn:lpt} using the \texttt{Python} code \texttt{velocileptors} \cite{Chen:2020ckc,Chen:2024pyp,Maus:2024dzi}\footnote{\url{https://github.com/sfschen/velocileptors}}
(see also~\cite{classpt,DAmico:2020kxu,Noriega:2024eyu,Linde:2024uzr,Harder:2026uei,Kobayashi:2026eee}). The one-loop power spectrum is described in terms of four bias parameters, $b_1, b_2, b_s, b_3$, as defined in Ref.~\cite{Chen:2020fxs}, where $b_3$ is the single, non-degenerate contribution from cubic bias operators at this order, which is related to the cubic tidal bias $b_{\Gamma_3}$
in the \texttt{class-pt} basis~\cite{Chudaykin:2020aoj}. We adopt the parametrization of Ref.~\cite{Maus:2024dzi}, scaling bias parameters of leading order $n$ by $r^n$, where $r = \sigma_8(z) / \sigma_8^{\rm fid}(z)$ and $\sigma_8^{\rm fid}$ is computed within the best-fit Planck $\Lambda$CDM cosmology \cite{Planck2018}, and we adopt Gaussian priors $\mathcal{N}(0,5^2)$ and $\mathcal{N}(0,1^2)$ for the cubic and quadratic bias parameters. For the linear bias, we transform to the Eulerian parameter, $b_1^E = 1 + b_1$, since it corresponds more directly to the physical amplitude of galaxy clustering, and use a uniform prior $\mathcal{U}[0,5]$.

We compute the tree-level bispectrum using the resummed linear power spectrum in Eq.~\ref{eqn:plin-res} and the quadratic galaxy kernels
\begin{align}
    Z_2(\vec k_1, \vec k_2) &= \frac{b_2^E}{2} + b_{\mathcal{G}_2}^E  \left( \frac{(\vec k_1 \cdot \vec k_2)^2}{k_1^2 k_2^2} - 1 \right) \nonumber \\
    &+ b_1 F_2(\vec k_1, \vec k_2) + f \mu^2 G_2(\vec k_1, \vec k_2) \nonumber \\
    &+ \frac{f\mu k}{2} \left( \frac{\mu_1}{k_1} Z_1(\vec k_1) + \frac{\mu_2}{k_2} Z_1(\vec k_1) \right) \nonumber 
\end{align}
where $\mu_n = \hat{k}_n \cdot \hat{n}$ and $\mu = \hat{k} \cdot \hat{n}$ for total momentum $\vec k = \vec k_1 + \vec k_2$ and the quadratic Eulerian biases are related to their Lagrangian counterparts as
\begin{equation}
    b_2^E = b_2 + \frac43 b_s, \quad b_{\mathcal{G}_2}^E = b_s - \frac27 b_1,
\end{equation}
such that the deterministic part of the tree-level bispectrum is given by cycling over the three momentum  $\vec k_{1,2,3}$ pairs
\begin{equation}
    B^{\rm PT}(\vec k_1, \vec k_2) = 2 Z_2(\vec k_1, \vec k_2) P^{\rm res}_{\rm lin}(\vec k_1) P^{\rm res}_{\rm lin}(\vec k_2) + \text{ cycl.}
\end{equation}
The resummed linear power spectrum for a given $\vec k$ is a function of $k, \mu$; since it cannot be expressed in a simple functional form outside of Eq.~\ref{eqn:plin-res}, we approximate it via a multipole expansion.
We find that retaining multipole moments up to hexadecapole provides numerical convergence well beyond the statistical precision of the data.

In addition to the dimensionless bias parameters, the EFT contains counterterms and stochastic parameters to renormalize the dependence on small, unresolved scales. We adopt the following counterterm for the power spectrum
\begin{equation}
    \delta_{\rm ctr}(\vec k) = \frac12 (\alpha_0 + \alpha_2 \mu^2 + \alpha_4 \mu^4) k^2 \delta_{\rm lin}(\vec k)
\end{equation}
with priors following \cite{desi1}.
For the stochastic power spectrum we use
\begin{equation}
    P_{\rm stoch}(\vec k) = \frac{1}{\bar{n}} \left( 1 + P_{\rm shot} + a_0 (k R_h)^2 + a_2 (k \sigma_v \mu)^2 \right)
\end{equation}
where $P_{\rm shot}$, $a_0$ and $a_2$ denote departures from the Poisson shot noise $\bar{n}^{-1}$, which is subtracted at the level of the data, and are sampled with prior $\mathcal{N}(0,1)$. We do not vary $a_0$ in our fits since its effect is set by the halo radius rather than halo velocity dispersion, and therefore suppressed relative to $a_2$ \cite{Schmittfull:2020trd,Chen:2020fxs,Ivanov:2024hgq,Ivanov:2024xgb,Chen2026}. The tree-level bispectrum also features stochastic contributions---we follow \cite{Ivanov:2021kcd,Philcox:2021kcw,MoradinezhadDizgah:2020whw,Oddo:2021iwq,Rizzo:2022lmh,Ivanov:2023qzb,desi1} and parametrize these via parameters $A_{\rm shot}, B_{\rm shot}$ describing the bispectrum and density dependence of the noise, marginalizing over them with unit Gaussian priors. Finally, the influence of the virial motions of galaxies inside halos, \textit{i.e.}\ the so-called Fingers of God, can lead to terms naively at higher order in perturbation contribute at one-loop order; in order to take into account these effects we include~\cite{Ivanov:2019pdj,Chudaykin:2020hbf,Ivanov:2021kcd}
\begin{align}
    \delta^P_{\rm FoG} &= c_{\rm FoG} (k\mu)^4 (b_1^E + f\mu^2) \delta_{\rm lin}(\vec k)\nonumber \\
    \delta^B_{\rm FoG} &= c_1 \ (k\mu)^2 \delta_{\rm lin}(\vec k)
\end{align}
in the power spectrum and bispectrum, respectively, again following the priors in \cite{desi1}.\footnote{For $c_{\rm FoG}$, we have adopted a more restrictive prior of $\mathcal{N}(-1, 0.5^2)$ compared to Tab.~III in the former work~\cite{desi1}.}

\subsection{Look-Elsewhere Effect}

\noindent Our constraints on primordial features span wide ranges of frequencies: since we do not know \textit{a priori} at which specific frequencies such feature may appear, for any tentative detection we need to account for the fact that more noise fluctuations exceeding a given significance threshold are expected when scanning over multiple frequencies rather than a single one. This phenomenon is known as the ``look-elsewhere effect'' \cite{Gross2010}. For example, for a discrete set of $N_{\rm trials}$ independent trials, the global $p$-value of a detection in a single trial of significance $p$ is $p_{\rm global} = N_{\rm trials} p_{\rm local}$, which we can equivalently phrase in terms of the global-to-local ratio, or trial factor, equal in this case to $N_{\rm trial}$. 

For continuous variables like the frequencies we study in this work, determining the trial factor is more involved. Roughly, one can think of the problem in terms of the statistics of upcrossings of a random variable describing the detection significance scanning over frequencies past a given threshold, with the excess number of upcrossings related to the global significance and the threshold corresponding to the local one \cite{Gross2010}. The corresponding statistical problem has been the subject of a wide literature ranging from analytic \cite{Davies1977,Davies1987} to simulation-based \cite{Gross2010,Bayer2020,Bayer2021}. 

In this work, we will estimate the look-elsewhere corrections for the oscillatory searches considered in this work, following the analytic calculations in Ref.~\cite{Davies1987} bounding the global $p$-value for a $\chi^2$ random variable, which exactly corresponds to a cosmic-variance limited search for oscillatory features with amplitude $A_X^2 = (A^X_{\rm sin})^2 + (A^X_{\rm cos})^2$ in the linear power spectrum. The details of this calculation are provided in Appendix~\ref{app:look-elsewhere}. Briefly, we obtain the trial factors
\begin{align}
    N_{\rm trials}^{\rm lin} &= 1 + \frac{k_{\rm max}}{k_\ast} \sqrt{\frac{3 u}{160\pi}} \Delta \omega_{\rm lin} \nonumber \\
    N_{\rm trials}^{\rm log} &= 1 +  \sqrt{ \frac{u}{18\pi} } \Delta \omega_{\rm log}
    \label{eqn:ntrial}
\end{align}
where $\Delta \omega_{\rm lin, \log}$ are the prior ranges of the respective parameters.
For $k_{\rm max} = 0.20\hMpc$ and our fiducial frequency ranges, this amounts to trial factors of roughly $50$ and $31$ for the linear and logarithmic oscillations at $3\sigma$, respectively. While we fit over a broader range of scales than $0.20\hMpc$ in practice, the above values of $N_{\rm trials}$ amount to a conservative choice since the oscillations are damped by IR displacements past that scale, resulting in weaker sensitivity to primordial features on nonlinear scales than is included in our linear theory forecast. Note that the logarithmic result does not explicitly depend on $k_{\rm max}$---instead, the dependence on fitting range is implicitly encoded in $\Delta \omega_{\rm log}$, since the lowest frequency in $\log k$ probed is determined by the ratio $k_{\rm max}/k_{\rm min}$.

We have chosen the analytic route since our MCMC runs do not sufficiently sample the parameter space to map out the upcrossings of the local significance across logarithmic or linear frequencies, in a way that would allow efficient and convergent utilization of the methods in e.g. Refs.~\cite{Gross2010,Bayer2020,Bayer2021}. This is sufficient for our purposes since we do not detect any features of unexpected significance and the broadband terms should be quite orthogonal to the oscillatory features we search for, though we have checked that our derived trial factors are in qualitative agreement with the numerical methods referenced above where possible.

\section{Cosmology Constraints from the Lagrangian EFT }
\label{sec:cosmo}

\noindent In order to set a baseline for our oscillatory feature analyses, we perform $\ld$ cosmology fits using the power spectrum and bispectrum measurements described above. This analysis differs from \citep{desi1} due to (a) the inclusion of the real-space power spectrum $Q_0$, (b) the Lagrangian forward model, and (c) the finer $k$-space binning and larger $k_{\rm max}$. 
In Appendix~\ref{app:pt_challenge}, we validate our analyses using the PT-challenge mock dataset \citep{Nishimichi:2020tvu}, finding that our LPT-based pipeline is well-suited to recover unbiased constraints from the DESI data (whose effective volume is far smaller than that of the simulation suite).

\begin{figure}
    \centering
    \includegraphics[width=\linewidth]{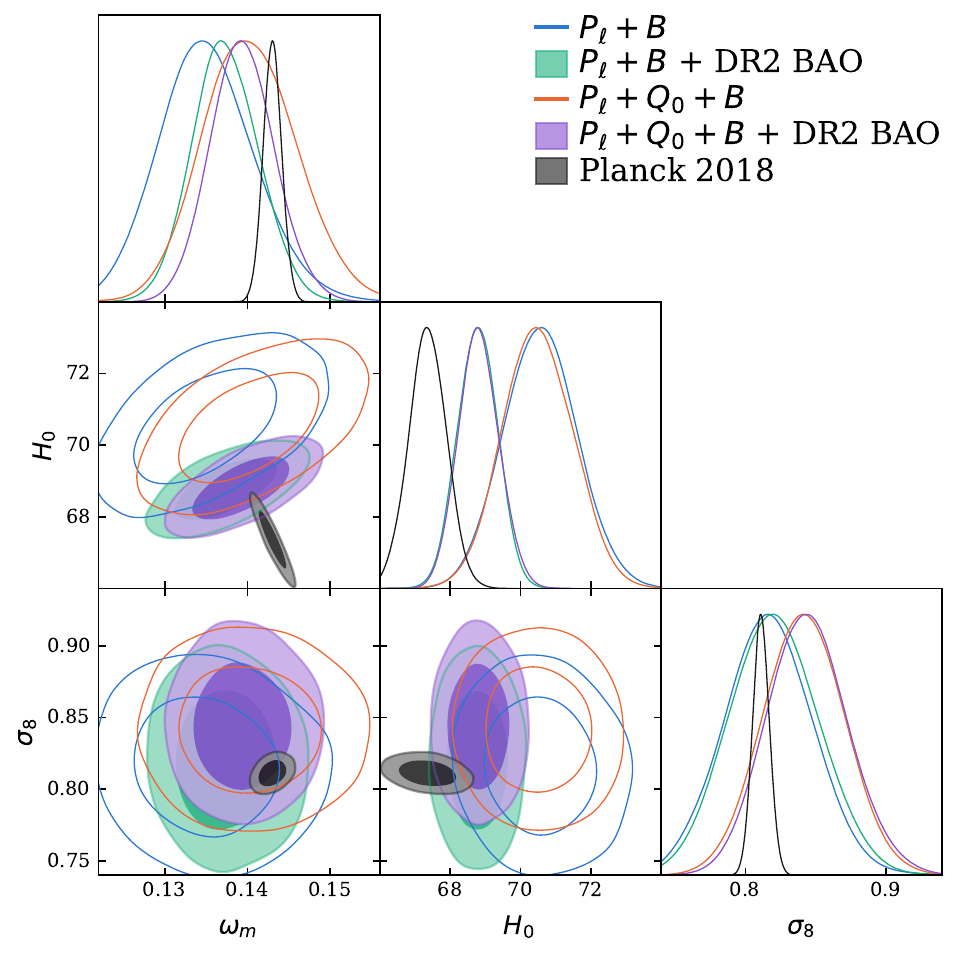}
    \caption{{Baseline $\Lambda$CDM constraints on $\omega_m$, $H_0$ and $\sigma_8$ from the DESI DR1 full-shape data, showing the 68\% and 95\% credible regions. We compare the redshift-space power spectrum and bispectrum analysis ($P_\ell+B$, blue) with that supplemented by the real-space power spectrum proxy $Q_0$ ($P_\ell+Q_0+B$, red). Results in filled (empty) contours show constraints obtained with (without) the DESI DR2 BAO likelihood. The \textit{Planck} 2018 posterior~\cite{Planck} is shown in dark gray for reference.}}
    \label{fig:lcdm_bao}
\end{figure}

Figure~\ref{fig:lcdm_bao} shows the $\Lambda$CDM constraints from the DESI data, derived from analyses both with and without $Q_0$, and with and without DESI DR2 BAO data. The addition of $Q_0$ results in two relatively small shifts in the constraints compared to the pure $P+B$ analysis: the matter density $\omega_m$ shifts slightly upwards, improving agreement with CMB data such as the \textit{Planck} 2018 cosmological constraints (shown in the figure), and the matter clustering amplitude $\sigma_8$ shifts slightly upwards. Adding in DR2 BAO primarily lowers the $H_0$ posterior, again bringing constraints into better agreement with the CMB. 

Despite the aforementioned differences, our results are consistent with those of the Eulerian EFT re-analysis of DESI full-shape data from~\cite{desi1}.
In particular, 
our $P_\ell+Q_0+B+\mathrm{DR2~BAO}$ combination yields $H_0=68.82\pm 0.58$ km/s/Mpc, $\Omega_m=0.2943\pm 0.0063$, $\sigma_8=0.844\pm 0.029$, which are consistent with the $P_\ell+B+$DR2~BAO results from~\cite{desi1}: $H_0=68.82\pm 0.58$ km/s/Mpc (identical to ours), $\Omega_m=0.2960\pm 0.0070$, $\sigma_8=0.818\pm 0.029$.
Our results also
agree well 
with DESI's official full-shape limits $H_0=68.56\pm 0.75$ km/s/Mpc, $\Omega_m=0.2962\pm 0.0095$, $\sigma_8=0.842\pm 0.034$, based on $P_\ell$+DR1 BAO~\cite{DESI:2024hhd}. 
Since $Q_0$
is the only new contribution to the data vector compared to~\cite{desi1}, we expect the improvement over that analysis to be driven 
primarily by this statistic. In agreement with that, our  
constraints 
on $\Omega_m$ are better than those of~\cite{desi1} by about $10\%$, which is indeed similar to  the improvement due to $Q_0$ previously seen on the BOSS data~\cite{Ivanov:2021fbu}. 
It is also worth mentioning that 
$Q_0$ ``pulls''
$\omega_m=\Omega_m h^2$ and $\Omega_m$ up by $0.5\sigma$,
making them more consistent with 
\textit{Planck}
CMB values.\footnote{For reference, our $P_\ell+B+Q_0+$DR2~BAO results are $\omega_m=0.1394\pm 0.0039$ ($\Omega_m=0.2943\pm 0.0063$), whilst 
the $P_\ell+B+$DR2~BAO constraint is $\omega_m=0.1374\pm 0.0040$ ($\Omega_m=0.2903\pm 0.0066$).}

Our main feature analyses simultaneously vary the underlying $\Lambda$CDM parameters with the oscillation amplitudes, resulting in cosmological posteriors when the primordial power spectrum is allowed to deviate from a simple power law. In order to better understand the dependence of searches for inflationary features on the background cosmology, we will also perform fits using the best-fit $\Lambda$CDM cosmology from the $P+B+Q_0$ chains excluding DR2 BAO, which amounts to the closest $\Lambda$CDM cosmology to our data, along with fits using the best-fit $\Lambda$CDM cosmology from the Planck analyses \cite{Planck2018}. In addition, given the tensions between these two datasets, particularly in terms of the expansion history, we will further derive constraints under the $w_0 w_a$CDM model, combining the DESI DR1 data with DESI DR2 BAO and supernovae from the Pantheon+ sample \cite{Brout:2022vxf}. We use the resulting best-fit $w_0 w_a$CDM parameters as a further template for our primordial features search. These constraints, as well as the $\Lambda$CDM ones discussed above, are listed in Table~\ref{tab:cosmologies}.

\begin{table*}[t]
    \centering
    \small
    \setlength{\tabcolsep}{8pt}
    \renewcommand{\arraystretch}{1.15}
    \resizebox{1.0\textwidth}{!}{%
    \begin{tabular}{llccccc}
        \toprule
        Analysis &
        Cosmology/feature &
        $H_0$ &
        $\omega_m$ &
        $n_s$ &
        $w_0$ &
        $w_a$ \\
        \midrule

        \textit{Planck} 2018
        & $\Lambda\mathrm{CDM}$
        & $67.36\pm0.54$
        & $0.1431\pm0.0013$
        & $0.9649\pm0.0042$
        & ---
        & --- \\

        \midrule

        DESI FS
        & $\Lambda\mathrm{CDM}$
        & $70.50\pm1.00$
        & $0.1403\pm0.0057$
        & $1.0208\pm0.0275$
        & ---
        & --- \\

        DESI FS + BAO + SNe
        & $w_0w_a\mathrm{CDM}$
        & $68.80\pm0.95$
        & $0.1452\pm0.0053$
        & $0.9933\pm0.0274$
        & $-0.892\pm0.060$
        & $-0.509\pm0.277$ \\

        DESI FS log-feature candidate
        & $(\omega_{\mathrm{log}},A_{\mathrm{log}})
           =(14.17,\,0.0238)$
        & $70.10$
        & $0.1445$
        & $0.9928$
        & ---
        & --- \\

        \bottomrule
    \end{tabular}
    }
    \caption{
        \textbf{Cosmological parameters used in the analysis.}
        The \textit{Planck} 2018 row reports the baseline
        TT,TE,EE+lowE+lensing posterior \citep{Planck}.
        The $\Lambda\mathrm{CDM}$ and $w_0w_a\mathrm{CDM}$ rows report
        the posterior means and $1\sigma$ uncertainties obtained using
        the DESI DR1 FS $P_\ell+Q_0+$bispectrum likelihood.
        The $w_0w_a\mathrm{CDM}$ analysis additionally includes
        Pantheon+ SNe and DESI DR2 BAO. In the feature analyses of Sec.~\ref{sec:results}, we fix the background cosmology to the best-fits from each of these analyses in turn, additionally performing joint fits of $\ld$ parameters and oscillatory amplitudes. The final row gives the maximum-likelihood
        $\Lambda\mathrm{CDM}$ cosmology at the
        $\omega_{\mathrm{log}}\simeq15$ log-feature candidate; because it
        corresponds to a single sample, no confidence interval is reported.
    }
    \label{tab:cosmologies}
\end{table*}

\section{Constraints on Inflationary Features}
\label{sec:results}

\noindent Given the above results, we conduct searches for primordial features across four configurations: (1) fixing cosmological parameters to the best-fit \textit{Planck} 2018 $\ld$ values; (2) fixing to the DESI DR1 $\ld$ values; (3) fixing to the best-fit $\wa$ parameters from DESI and supernova data; (4) allowing the $\ld$ parameters to float together with parameters describing the primordial features.

\begin{figure}
    \centering
    \includegraphics[width=\linewidth]{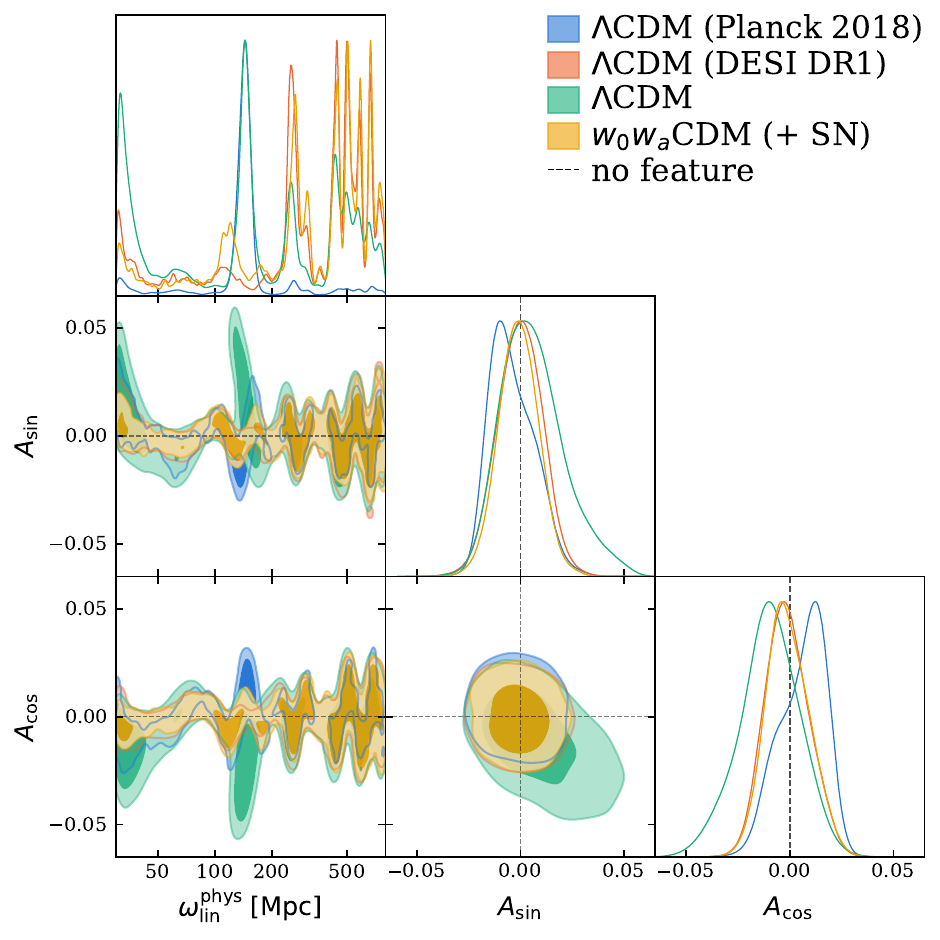}
    \caption{Posterior distributions of the linear-oscillation parameters, the frequency $\omega^{\rm phys}_{\rm lin} = \omega_{\rm lin} / k_\ast$ and the sine and cosine amplitudes $A_{\rm sin}$ and $A_{\rm cos}$, from the DESI DR1 power spectrum and bispectrum analysis. The four chains correspond to the cosmology being fixed to the \textit{Planck} 2018 (blue) or DESI DR1 (orange) $\Lambda$CDM best fits, to the $w_0w_a$CDM best fit from DESI+BAO+SN (yellow), and to $\Lambda$CDM with all cosmological parameters varied (green). Dashed lines mark the feature-free model. The isolated peak of the fixed-cosmology chains at $\omega_{\rm lin}^{\rm phys}\simeq140~{\rm Mpc}$ coincides with the BAO scale and disappears when the cosmology is free.}
    \label{fig:linear_triangle}
\end{figure}

\begin{figure}[htb!]
    \centering
    \includegraphics[width=\linewidth]{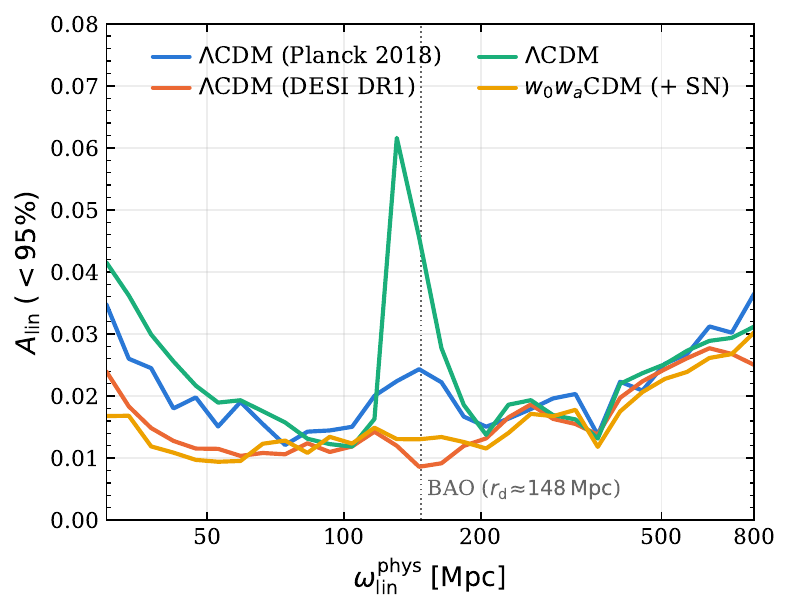}
    \caption{Upper bounds (95\% CL) on the total amplitude $A_{\rm lin}=\sqrt{A_{\rm sin}^2+A_{\rm cos}^2}$ of linear oscillations as a function of frequency $\omega^{\rm phys}_{\rm lin} = \omega_{\rm lin}/k_\ast$, for the four analysis configurations of Figure~\ref{fig:linear_triangle}. The vertical dotted line marks the BAO scale, $r_d\simeq148~{\rm Mpc}$. The bounds are at the $1$--$2\%$ level over most of the range; the free-$\Lambda$CDM analysis loses constraining power around the BAO scale, where the feature is degenerate with the acoustic scale, and at the lowest frequencies, where it is degenerate with the broadband shape.}
    \label{fig:linear_constraints}
\end{figure}

Figs.~\ref{fig:linear_triangle} and~\ref{fig:linear_constraints} show our results for linear oscillations. The former plot shows the posteriors on these features from our MCMC chains, scanning over feature frequencies and amplitudes: in all four scenarios there are many $1-2\sigma$ excursions in the posterior for primordial features; this can be expected given the look-elsewhere effect. However, when cosmology is fixed to the \textit{Planck} 2018 best-fit, we find a $\Delta \chi^2 = 9.8$ preference for a feature with $\omega_{\rm lin}^{\rm phys} = 140\,\text{Mpc}$ and an amplitude $A_{\rm lin} = 1.3\%$. This feature is coincident with the BAO scale and, moreover, around $50\%$ and $30\%$ of the $\Delta \chi^2$ comes from the LRG2 and LRG1 samples, respectively. This suggests that its origin is related to the distance-scale tension between the CMB and DESI data, which are also primarily located in those bins~\cite{Wang:2024pui,Liu:2024gfy,Chudaykin:2024gol}. Conversely, no such candidate emerges when the $\Lambda$CDM parameters are freed; instead, when the BAO feature is allowed to be cosmology dependent, our constraints at the approximate BAO scale weaken. In Figure~\ref{fig:log_constraints} we show the corresponding $95\%$ bounds on the total feature amplitude $A_{\rm lin}$, which we obtain as a function of frequency by binning our MCMC chains over narrow frequency ranges and computing the resulting 1D amplitude posteriors. As can be seen in the figure, the $95\%$ bounds on the total feature amplitude weaken from an amplitude of $1-2\%$ at most surveyed frequencies to around $6\%$ at the BAO frequency; moreover, this region occupies most ($57\%$) of the posterior volume. This occurs because variations in the cosmological parameters shift the position of the BAO peak, leading to a degeneracy with linear primordial oscillations of frequency $\omega_{\rm lin}^{\rm phys}\simeq r_s$ \cite{Beutler:2019ojk,Ballardini:2019tuc}. At the largest scales, the free $\Lambda$CDM constraints also weaken rapidly, owing to the degeneracy of low frequency oscillations with cosmological parameters constrained by the overall power spectrum shape and the finite range of scales probed by the data.

In total, we find that the linear oscillation amplitude is constrained at $95\%$ confidence to
\begin{align}
    A_{\rm lin} < 0.024 \text{ (\textit{Planck} 2018)} \nonumber \\
    A_{\rm lin} < 0.033 \text{ (free }\Lambda\text{CDM)},
\end{align}
marginalizing over the phase and frequency.
The other two fixed-cosmology analyses behave similarly to those with the \textit{Planck} 2018 parameters, with the best constraints coming from the $w_0 w_a$CDM variant with $A_{\rm lin} < 0.021$. If the BAO feature is excluded by removing frequencies within $\pm 20\%$ of the feature, the free-$\Lambda$CDM chain instead recovers $A_{\rm lin} < 0.027$.

\begin{figure}
    \centering
    \includegraphics[width=\linewidth]{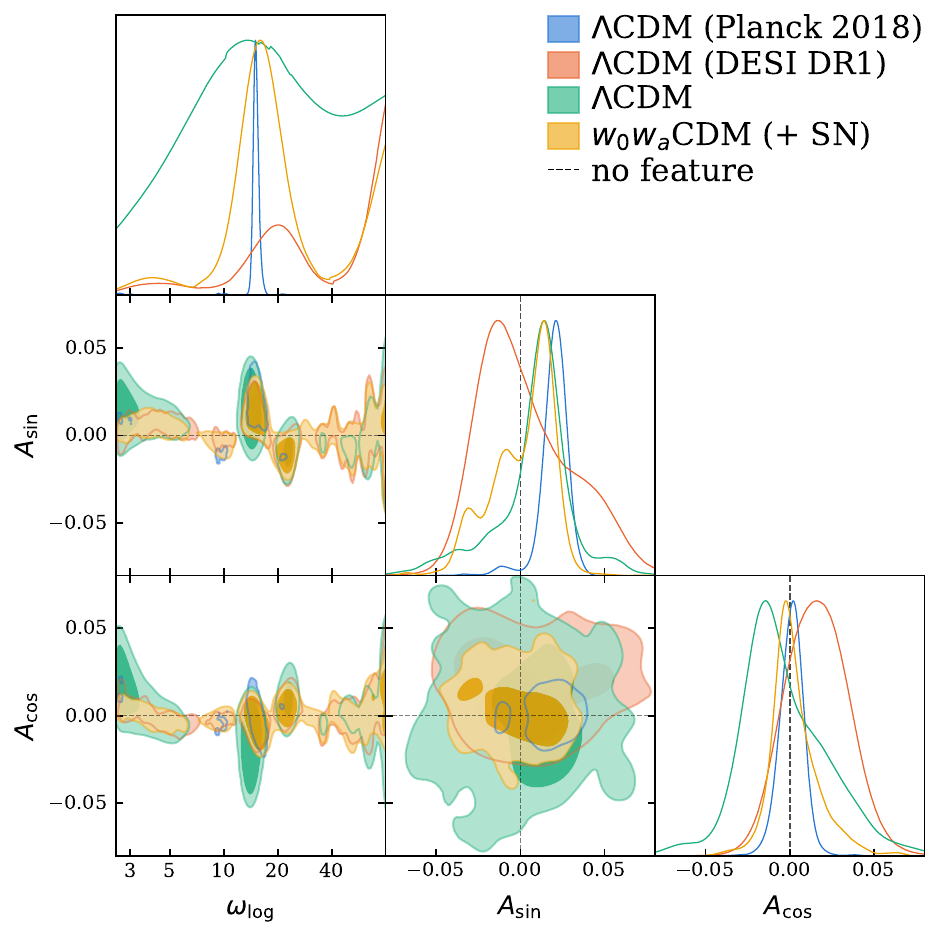}
    \caption{As Figure~\ref{fig:linear_triangle}, but for logarithmic oscillations, parameterized by the frequency $\omega_{\rm log}$ and the amplitudes $A_{\rm sin}$, $A_{\rm cos}$. All chains except the free-$\Lambda$CDM variant show a peak at $\omega_{\rm log}\simeq15$, which corresponds to a local linear frequency close to the BAO scale at $k\simeq0.1 \text{Mpc}^{-1}$; it is strongest for the fixed \textit{Planck} 2018 cosmology.}
    \label{fig:log_triangle}
\end{figure}

\begin{figure}
    \centering
    \includegraphics[width=\linewidth]{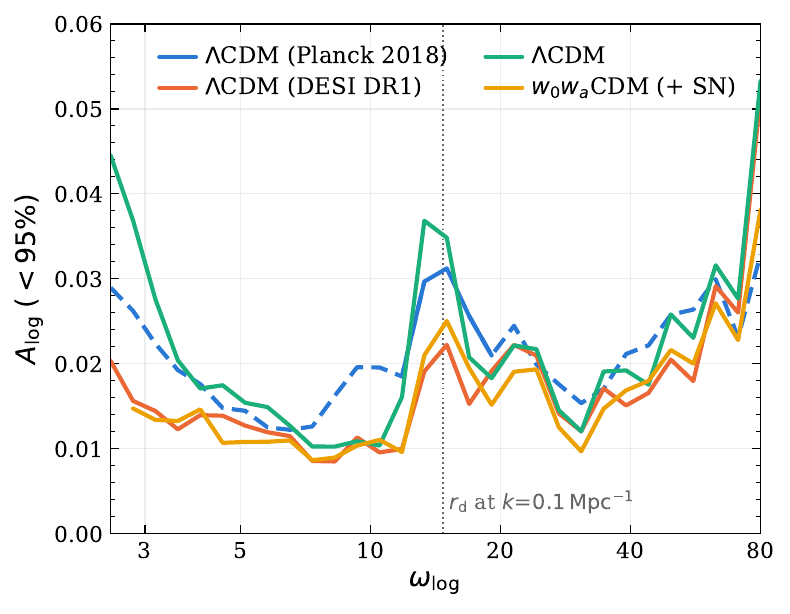}
    \caption{Upper bounds (95\% CL) on the amplitude $A_{\rm log}$ of logarithmic oscillations as a function of $\omega_{\rm log}$ for the four analysis configurations of Figure~\ref{fig:log_triangle}. The vertical dotted line marks the frequency at which the local wavelength matches the sound horizon $r_d$ at $k=0.1~{\rm Mpc}^{-1}$. The bounds reach the $1\%$ level for $\omega_{\rm log}\in(5,10)$ and degrade around $\omega_{\rm log}\simeq15$, where the feature mimics the BAO, and at both ends of the range. We run separate chains bracketing the BAO scale at $\omega_{\rm log} \sim 15$, denoted by the blue dashed lines, to improve convergence of bounds in regions where the original chain (solid blue) has little posterior volume.}
    \label{fig:log_constraints}
\end{figure}

Figs.~\ref{fig:log_triangle} and~\ref{fig:log_constraints} similarly show the posteriors and per-frequency $95\%$ bounds on logarithmic features. Our chains return uncomplicated posterior bounds on these features, with no $>2\sigma$ excursions except for $\omega_{\rm log} \approx 15$, where we can clearly see peaks in the posterior for $\omega_{\rm log}$ and a sharp degradation in the constraining power on the amplitude. Intriguingly, this feature corresponds precisely to a local \textit{linear} frequency of $15 / k = 150\ k_{0.1}^{-1}\ \text{Mpc}$ at $k = 0.1\ \text{Mpc}$, or $k = 0.15 \hMpc$ in $h$-units, near the scale where DESI constrains baryon acoustic oscillations best, with logarithmic oscillations at around the $2\%$ level. The strongest detection arises in the fixed \textit{Planck} cosmology chain, at $\Delta \chi^2 = 18.5$, corresponding to a local significance of $3.9\sigma$, or a global significance just shy of $3\sigma$ after accounting for the look-elsewhere effect. Moreover, the best-fit including logarithmic oscillations within the \textit{Planck} 2018 cosmology has a $\chi^2$ very similar to that of the fits without \textit{Planck} or features, with again around $64\%$ of the improvement coming from the LRG2 sample. As before, this is highly suggestive of being related to the tension between the CMB and DESI BAO data within $\Lambda$CDM. The fixed $w_0 w_a$CDM chain exhibits the same feature at a local significance of $2.5\sigma$, while the DESI only $\Lambda$CDM chains show no significant detection at all, providing further support for this interpretation.

\begin{figure}
    \centering
    \includegraphics[width=\linewidth]{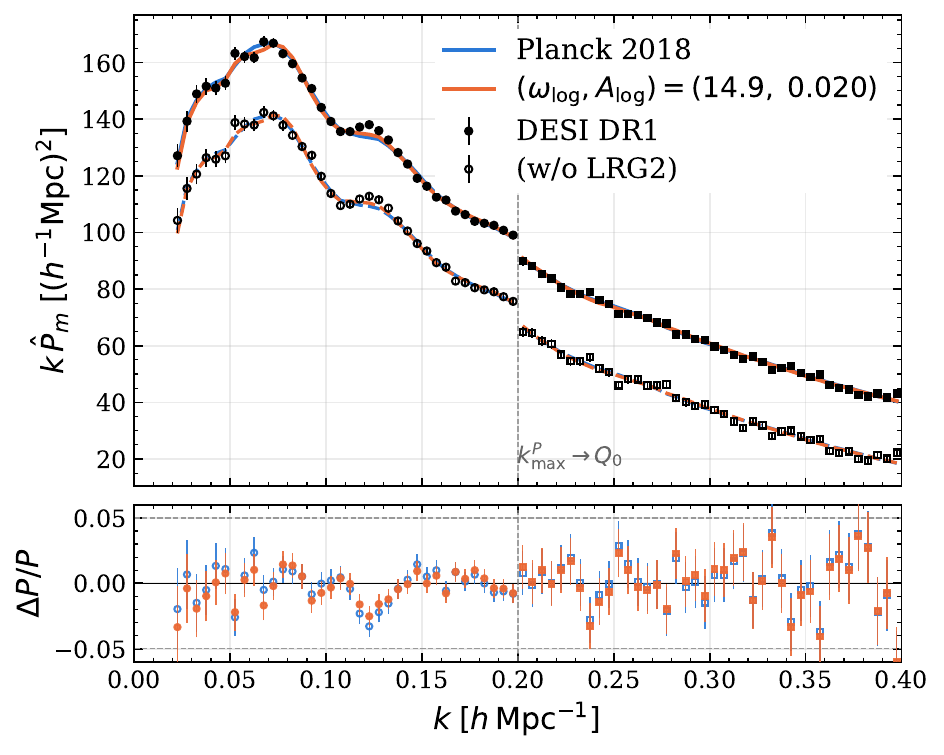}
    \caption{Reconstructed real-space matter power spectrum obtained by combining the DESI DR1 samples (black points; open points exclude the LRG2 sample) with the inverse-variance weights of Eq.~\eqref{eqn:synth}, compared with the best-fit \textit{Planck} 2018 predictions without (blue) and with (orange) the logarithmic feature $(\omega_{\rm log},A_{\rm log})=(14.9,\,0.020)$. The two sets of points are offset vertically for clarity. The vertical dashed line marks the transition from the $P_\ell$ to the $Q_0$ data at $k_{\rm max}=0.2\hMpc$. The lower panel shows the residuals with respect to the feature-free model. The excess around the second BAO peak, $k\simeq0.12\hMpc$, is driven by the LRG2 sample.}
    \label{fig:synthesized_mPk}
\end{figure}

To explore the putative primordial feature candidate further, Figure~\ref{fig:synthesized_mPk} shows the ``reconstructed'' matter power spectrum across the DESI samples compared to the best-fit prediction in the \textit{Planck} 2018 cosmology, with and without logarithmic oscillations. Specifically, we combine
\begin{equation}
    \hat{P}_m^i = \frac{1}{D(z_i)^2 (b_i^2 + \frac23 b_i f(z_i) + \frac15 f(z_i)^2)} \hat{P}_0
    \label{eqn:synth}
\end{equation}
for each sample, inverse-variance weighting by the effective volumes for each tabulated in Ref.~\cite{DESI:2024aax},
where $b_i,D(z_i),f(z_i)$
are the linear bias, 
growth factor and logarithmic growth factor of a tracer $i$
and $\hat{P}_0$ is the monopole data vector. For $k>0.2\hMpc$, we instead show $Q_0/(D(z_i)b_i)^2$, using that $Q_0$ is equal to the  real-space galaxy power spectrum in the limit that the redshift-space power spectrum is well-approximated by its $\ell = 0, 2, 4$ multipoles. Note, however, that the two matter power spectrum reconstructions do not coincide exactly at $k = 0.2 \hMpc$ due to the effect of nonlinearities.
In any case, the \textit{Planck} 2018 cosmology evidently provides a fairly poor fit to the data around the second BAO peak, around which $\omega_{\rm log} = 15$ approximates the apparent BAO frequency better. Indeed, removing the LRG2 sample visibly reduces this discrepancy.

Over the total range we cover, our results bound logarithmic inflationary features at
\begin{align}
    A_{\rm log} < 0.031 \text{ (\textit{Planck} 2018)} \nonumber \\
    A_{\rm log} < 0.035 \text{ (Free }\Lambda\text{CDM)}.
\end{align}
However, these bounds under-state our constraints in most frequency ranges. As can be seen in Figure~\ref{fig:log_constraints}, we constrain logarithmic features to the $1\%$ level for $\omega_{\rm log} \in (5,10)$ and at the $2\%$ level for $\omega_{\rm log} \in (20, 40)$ for all model configurations. This reflects that we lose constraining power at large frequencies due to the finite window of the survey, and at small frequencies due to degeneracies with cosmological parameters. If we exclude the BAO region $\omega_{\rm log} \in (10,20)$, where primordial features are presumably degenerate with the BAO signal, we further obtain $A_{\rm log} < 0.021$ from the \textit{Planck} 2018 chains, a $50\%$ stronger bound that is comparable to those from the fixed-cosmology DESI $\Lambda$CDM and $w_0 w_a$CDM chains.

\section{Discussion and Conclusions }
\label{sec:conclusions}

\noindent In this work we have searched for oscillatory features in the 
primordial power spectrum using the two- and three-point functions of redshift-space galaxy and quasar clustering from the public DESI DR1 data.
Our analysis has several new ingredients with respect to previous LSS searches. 
First, we consistently include the tree-level galaxy bispectrum,
which sharpens the constraints on the standard cosmological and nuisance parameters and thereby reduces 
their degeneracy with the feature parameters. 
Second, we use the Lagrangian effective field theory, which allows us to
model the non-linear damping of the features due to IR displacements 
in a principled way without 
splitting it into smooth and oscillatory parts. This is particularly important for our case as we consider
linear 
power spectra with both the BAO and primordial features.
We have validated our LPT pipeline 
on the PT Challenge simulation (Appendix~\ref{app:pt_challenge}), which includes the LPT analysis of the bispectrum for the first time. 
Third, we have studied the dependence of the feature searches on the assumed 
background cosmology.
We explored multiple options: fixing cosmology to the \textit{Planck} 2018 and DESI DR1 $\Lambda$CDM best fits or the $w_0w_a$CDM best fit from DESI+BAO+SNe, as well as fully marginalizing over $\Lambda$CDM parameters in a CMB-independent fit to DESI FS and BAO data. 
Finally, we have estimated the look-elsewhere effect analytically.

Our main results are as follows. For linear oscillations, we find no significant 
evidence for a feature once the cosmological parameters are varied, 
and bound the amplitude to $A_{\rm lin}<0.033$ at 95\% CL, improving 
to $A_{\rm lin}<0.027$ when the frequencies within $20\%$ of the BAO scale are excluded. 
Over most of the frequency range the per-frequency bounds are at the $1$--$2\%$ level.
For logarithmic oscillations, the marginalized bound is $A_{\rm log}<0.035$, 
reaching the $1\%$ level for $\omega_{\rm log}\in(5,10)$ and the $2\%$ level for $\omega_{\rm log}\in(20,40)$.

When the cosmology is fixed to the \textit{Planck} 2018 best fit, the 
bounds tighten to $A_{\rm lin}<0.024$ and $A_{\rm log}<0.031$, but the 
chains also single out a logarithmic feature with $\omega_{\rm log}\simeq 15$ 
and $A_{\rm log}\simeq 2\%$ at $\Delta\chi^2=18.5$, \textit{i.e.}\ $3.9\sigma$ local and 
roughly $3\sigma$ global significance, as suggested by our look-elsewhere effect estimate. A weaker version of the same 
candidate, with a linear frequency close to the BAO scale, appears in the fixed-cosmology linear frequency features' chains.

We do not interpret this candidate as evidence for primordial physics. Its best-fitting wavelength coincides with the sound horizon at the wavenumbers 
where DESI constrains the BAO best, and the improvement in $\chi^2$ is dominated 
by the LRG2 and LRG1 samples, which are known to be
in tension with the baseline \textit{Planck} $\ld$ cosmology, see e.g.~\cite{DESI:2024mwx,Wang:2024pui,Liu:2024gfy,Chudaykin:2024gol}.
In addition, the best fit with the feature in the 
\textit{Planck} cosmology has essentially the same $\chi^2$ as the feature-free 
best fit in the DESI cosmology. This preference disappears both when the $\Lambda$CDM parameters are
varied or when they are fixed to the DESI DR1 best-fit. 

The most likely explanation for the anomalous feature is therefore that it absorbs the mismatch between the acoustic scale preferred by the DESI full-shape data 
and that of the \textit{Planck} 
$\Lambda$CDM cosmology, 
\textit{i.e.}\ the same 
distance-scale tension that has been interpreted by DESI as evidence for 
evolving dark energy~\cite{DESI:2025gwf} (and also curvature, negative neutrino masses, topology, and a host of other new physics proposals \citep[e.g.,][]{desi2,Chudaykin:2026amr,Green:2024xbb,Philcox:2025faf}).
This clearly illustrates
that searches for primordial features with fixed cosmology can be biased by even mild tensions in the background expansion history, and that marginalization over the cosmological parameters is essential for robust constraints, 
even though it weakens the bounds around the BAO frequency.

Our bounds are competitive with the CMB limits
$A_{\rm lin}<0.021$,
$A_{\rm log}<0.022$
from \textit{Planck} and ACT~\cite{Nerval:2026iev} (see also~\cite{Planck:2018jri}).
Specifically, away from the candidate
frequency 
our $A_{\rm lin}$
limit is $22\%$ worse
than the CMB one, 
but our limit on $A_{\rm log}$
for $\omega_{\rm log}<10$ is about $50\%$ better. Our result for $\omega_{\rm log}\in (20,40)$ is
similar to the CMB one.

Our constraints are somewhat weaker 
than 
the  
BOSS-based LSS constraints of~\cite{Beutler:2019ojk}, which found $\lesssim 1\%$
limits on both linear and logarithmic features.
We believe that this is primarily a result of the
marginalization over
a larger set of
nuisance parameters and cosmology in our analysis 
relative to that of~\cite{Beutler:2019ojk}. In addition, 
our result is based on the pre-reconstructed power spectra, which feature a larger suppression of the features than the 
post-reconstructed 
power spectra used by~\cite{Beutler:2019ojk}. While reconstruction can improve the signal-to-noise of primordial features, we do not consider it here since accurately modeling post-reconstruction oscillations signals would require further development of EFT predictions for the reconstructed field (but see e.g.~\cite{Hikage2017,Hikage2020,Chen2024,Sugiyama2024}).

Finally, our beyond-candidate limits 
are somewhat stronger
than the eBOSS results
$A_{\rm lin}<0.031$,
$A_{\rm log}<0.024$~\cite{Calderon:2025xod}. However, their joint LSS+CMB analysis quotes 
$A_{\rm lin}<0.022$,
which is 
$18\%$ stronger
than our limit; and
$A_{\rm log}<0.021$,
which is similar to our constraints
in the range 
$\omega_{\rm log}\in (20,40)$.

Our analysis can be extended in several directions. 
The one-loop bispectrum likelihood for DESI DR1 \citep{desi4} 
would facilitate extending the bispectrum to smaller scales, which is expected to yield additional constraints on features in the primordial power spectrum \cite{Euclid:2023shr,Karagiannis:2026xkv}, but this requires revisiting the current modeling
of the non-linear damping. 
Simulation-based priors on the EFT parameters \citep{desi5} and the 
addition of projected galaxy clustering information~(\cite{Chen:2024vuf,Maus:2025rvz,desi4}) may further tighten the bounds. 
Finally, the same pipeline can be applied to other features, such as 
the oscillating signatures 
of the cosmological collider in the galaxy power spectrum shape due to the scale-dependent bias~\cite{Cabass:2024wob}.
It will also be interesting 
to extend our analysis 
to the resonant features in the axion monodromy
inflation model, which predicts 
different, but correlated, oscillation patterns in the power spectrum and bispectrum~\cite{Flauger:2009ab,Flauger:2010ja}.
We leave this for future work.

\begin{acknowledgments}
{\small
\noindent Support for this work was provided by NASA
through the NASA Hubble Fellowship grant HST-HF2-51572.001 awarded by the Space Telescope
Science Institute, which is operated by the Association of Universities for Research in Astronomy,
Inc., for NASA, under contract NAS5-26555. 
AC acknowledges funding from the Swiss National Science Foundation.
OHEP is supported in part by the National Science Foundation under grant number PHY-2609396.
MMI is supported by the Office of High Energy Physics of U.S. Department of Energy under grant Contract Number  DE-SC0012567. 
}
\end{acknowledgments}

\appendix

\section{Validation on PT-Challenge Data}
\label{app:pt_challenge}
\begin{figure}
    \centering
    \includegraphics[width=\linewidth]{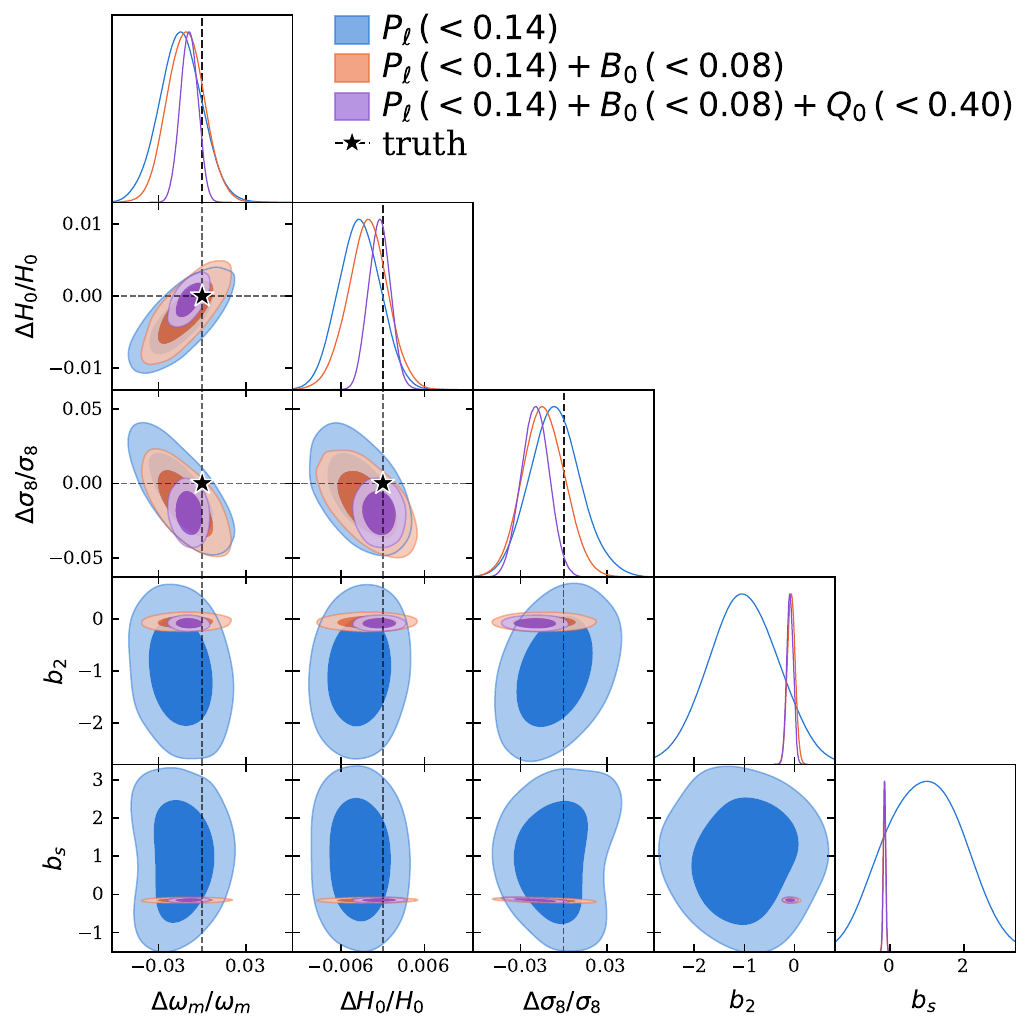}
    \caption{{Validation of the LPT-based pipeline on the PT Challenge simulation~\cite{Nishimichi:2020tvu}: recovered fractional shifts of $\omega_m$, $H_0$ and $\sigma_8$  with respect to the still masked true values (dashed lines and star), together with the Lagrangian bias parameters $b_2$ and $b_s$, from the redshift-space power spectrum alone with $k^P_{\rm max}=0.14\hMpc$ (blue) and from the power spectrum plus the bispectrum with $k_{\rm max}^B=0.08\hMpc$ (orange). Contours show the 68\% and 95\% credible regions.}}
    \label{fig:pt_challenge}
\end{figure}

\noindent In order to validate the combination of the power spectrum and bispectrum in LPT, we use our pipeline to fit data from the blinded PT challenge mocks \cite{Nishimichi:2020tvu}, which have previously been used to validate power spectrum modeling in LPT \cite{Chen:2020zjt}.\footnote{\url{https://www2.yukawa.kyoto-u.ac.jp/~takahiro.nishimichi/data/PTchallenge/}} These are ten $3072^3$ N-body simulations each with volume $(3.84\ \text{Gpc}/h)^3$, amounting to more than $20\times$ the effective volume of DESI DR1. Using the full covariance of these data, computed within the disconnected Gaussian limit, therefore acts as a stringent test on the validity of the model. Following the setup of the PT challenge, we sample over all $\Lambda$CDM parameters, which we report only in terms of their relative deviations from the blind true values, keeping only the baryon density $\omega_b$ fixed to its true value. In keeping with previous works fitting these mocks, we fit the one-loop power spectrum to a relatively large scale $< 0.14 \hMpc$, beyond which two-loop effects become comparable to the statistical error\footnote{Though we note that the resulting bias is still well below the statistical error of realistic galaxy surveys out to $k = 0.2 \hMpc$, as shown in Refs.~\cite{Nishimichi:2020tvu,Chen:2020zjt,Chudaykin:2024wlw}.} while fitting the tree level bispectrum and real-space power spectrum proxy $Q_0$ with the same fitting ranges used in the main analysis. As can be seen in Figure~\ref{fig:pt_challenge}, our LPT pipeline successfully recovers the correct cosmology to within $2 \sigma$ when fitting only the power spectrum, including the bispectrum, and including $Q_0$, while the addition of the bispectrum significantly tightens constraints on the quadratic bias parameters, as expected.

\section{Look-Elsewhere Effect}
\label{app:look-elsewhere}

In this Appendix we provide the details of the derivation of the look-elsewhere effect correction for linear and logarithmic features in Equation~\ref{eqn:ntrial}, following ref.~\cite{Davies1987}. Let us define the deviation of the observed power spectrum, $\hat{P}$, from a no-feature template power spectrum $P_t$, via $\hat{p}_i \equiv \hat{P}(k_i) / P_t(k_i) - 1$ for bin $i$. Given a feature with dispersion $f_\omega(k)$, \textit{i.e.}\ $\omega_{\rm lin}\,k/k_\ast$ or $\omega_{\rm log} \log(k)$, we can then define the optimal inverse-variance-weighted quantities
\begin{align}
    Z_c(\omega) &= \sum_i w_i \cos(f_\omega(k_i)) \hat{p}_i, \nonumber \\
    Z_s(\omega) &= \sum_i w_i \sin(f_\omega(k_i)) \hat{p}_i
\end{align}
with $w_i \propto \sigma^{-2}_i$, where $\sigma^2_i$ is the variance of $\hat{P}(k_i)$. In our simplified setup, the local significance of a feature with dispersion $f_\omega(k)$ is given by the random variable $Z^2 = Z_c^2 + Z_s^2$.

In the many-oscillation limit where $f_\omega$ runs over a range $\gg 2\pi$ across the scales probed by the analysis, one can show using double-angle formulas that $Z_{c,s}$ are uncorrelated at fixed $\omega$, \textit{i.e.}\ $Z^2$ constitutes a $\chi^2$ random variable. We can similarly show that
\begin{align}
    & \langle Z_c Z_c'\rangle = \langle Z_s Z_s'\rangle = 0 \nonumber \\
    & \langle Z_c Z_s'\rangle = -\langle Z_s Z_c'\rangle = \langle \partial_\omega f_\omega(k) \rangle_k \nonumber \\
    &  \langle Z_c' Z_c'\rangle = \langle Z_s' Z_s'\rangle = \langle (\partial_\omega f_\omega(k))^2 \rangle_k \nonumber \\
    & \langle Z_c' Z_s'\rangle = 0,
\end{align}
for $Z,Z'\equiv Z(\omega), \partial_\omega Z(\omega')$, where $\langle \cdots \rangle_k$ indicates the weighted average over $k$ modes, \textit{i.e.}\ over the weights $w_i$. Ref.~\cite{Davies1987} show that, when these conditions are satisfied, the global $p$-value is bounded by
\begin{equation}
    p_{\rm global}(u) \leq e^{-u/2} \left(1 + \frac{\sqrt u}{\pi}  \int_{\omega_{\rm min}}^{\omega_{\rm max}} E[ |\eta(\omega)| ] d\omega \right)
\end{equation}
where $\eta = (\eta_c, \eta_s)$ are independent random variables with variance $\langle (\partial_\omega f_\omega)^2 \rangle_{c,k}$. 

Since the local $p$-value is $e^{-u/2}$, the trial factor is therefore
\begin{align}
    N_{\rm trials} &\leq 1 + \frac{\sqrt u}{\pi}  \int_{\omega_{\rm min}}^{\omega_{\rm max}} E[ |\eta(\omega)| ] d\omega \nonumber \\
    &= 1 + \sqrt{\frac{u}{2\pi}} \int_{\omega_{\rm min}}^{\omega_{\rm max}} \sqrt{\langle (\partial_\omega f_\omega)^2 \rangle_{c,k}}\,d\omega.
\end{align}
For linear and logarithmic oscillations, we have
\begin{align}
    \partial_\omega f_\omega &= \frac{|k|}{k_\ast}, \; \rightarrow \langle (\partial_\omega f_\omega)^2 \rangle_{c,k} = \frac{3}{80} \left( \frac{k_{\rm max}}{k_\ast} \right)^2, \nonumber \\
    \partial_\omega f_\omega &= \log \frac{|k|}{k_\ast}, \; \rightarrow \langle (\partial_\omega f_\omega)^2 \rangle_{c,k} = \frac19,
\end{align}
where we have taken the $k_{\rm min} \rightarrow 0$ limit in both cases, \textit{i.e.}\ the trial numbers are
\begin{align}
    N_{\rm trials}^{\rm lin} &= 1 + \frac{k_{\rm max}}{k_\ast} \sqrt{\frac{3 u}{160\pi}} \Delta \omega_{\rm lin} \nonumber \\
    N_{\rm trials}^{\rm log} &= 1 +  \sqrt{ \frac{u}{18\pi} } \Delta \omega_{\rm log}
\end{align}
where $\Delta \omega_{\rm lin, \log}$ are the prior ranges of the respective parameters.

\bibliography{main_clean.bib}


\end{document}